# Ge$_{0.95}$Sn$_{0.05}$ on Si avalanche photodiode with Spectral Response Cutoff at 2.14 µm

JUSTIN RUDIE,[1,2] XIAOXIN WANG,[3] RAJESH KUMAR,[2] GREY ABERNATHY,[1,2,4] SYLVESTER AMOAH,[2] STEVEN AKWABLI,[2] HRYHORII STANCHU,[4] PERRY C. GRANT,[5] BAOHUA LI,[5] WEI DU,[1,2,4] JIFENG LIU,[4*], SHUI-QING YU[1,2,4,*]

[1]*Material Science and Engineering, University of Arkansas, Fayetteville, AR 72701 USA*
[2]*Department of Electrical Engineering and Computer Science, University of Arkansas, Fayetteville, AR 72701 USA*
[3]*Thayer School of Engineering, Dartmouth College, Hanover, NH, 03755 USA*
[4]*Institute for Nanoscience and Engineering, University of Arkansas, Fayetteville, AR 72701 USA*
[5]*Arktonics, LLC, 1339 S. Pinnacle Dr., Fayetteville, AR, 72701 USA*
*\*jifeng.liu@dartmouth,edu; syu@uark.edu*

**Abstract:** GeSn-based avalanche photodiode (APD) operating in shortwave infrared (SWIR) wavelength was demonstrated in this work. A separate absorption and charge multiplication (SACM) structure was employed to take advantage of long wavelength absorption in GeSn and low impact ionization ratio of Si. Due to lattice mismatch between Si and GeSn that would degrade GeSn material quality if with direct growth, a 240-nm-thick Ge buffer was utilized which simultaneously allows for the transporting photo generated electrons from GeSn absorber to Si multiplication layer. Spectral response showed the cut off wavelength beyond 2.1 µm at room temperature. Dart current and capacitance-voltage measurements indicated a punch-through voltage of -10 V. The measured responsivities were 0.55 A/W and 0.34 A/W under 1.55 µm and 1.9 µm excitation lasers, respectively. The maximum gain was obtained as 3.44 at 77 K under 1.9 µm laser. Even at 250 K, the calculated gain was greater than unity. Simulation of electric field distribution revealed that the GeSn is partially depleted at operating voltages, which can be improved by reducing the background doping levels in GeSn absorber and Ge buffer layer.

## 1. Introduction

Infrared (IR) Avalanche photodiodes (APDs) featuring internal gain caused by avalanche multiplication have been widely used in various applications where high sensitivities are needed, such as light detection and ranging (LiDAR) systems, fiber-optic telecommunications, eye-safe imaging, medical sensing, etc. [1-5]. It is well documented that the best material candidate for high performance APDs is Si, whose *k* factor (ratio of the electron and hole ionization coefficients) is as low as ~0.02 and gain-bandwidth product is as high as greater than 340 GHz [6, 7]. However, the bandgap energy of 1.12 eV prevents Si from operating at wavelengths longer than 1.1 µm. For the past two decades, there have been many efforts to address excess noise level and the gain-bandwidth product. A remarkable breakthrough has been made by employing a separate absorption-charge-multiplication (SACM) structure, which separates the absorption and multiplication regions by a charge layer, allowing for using excellent IR materials for high sensitivity light absorption and Si for impact ionization with a low excess noise factor. Promising results were obtained from Ge on Si APD, showing high sensitivity as high as those of the III-V compound APDs [7-9]. The limitations of such APDs are mainly from i) the lattice mismatch between Ge and Si leads to high dark current that offsets the low excess noise factor; and ii) spectral response cut-off at 1.6 µm. AlInAsSb on GaSb and InGaAs on Si APDs have been demonstrated and exhibit high performance at longer IR wavelengths [10-13]. Nevertheless, challenges of these APDs lie in high expense of III–V epitaxial wafers and difficulties of integrating III-V materials into current Si fabrication facility. Therefore, developing an IR APD based on a monolithic, complementary metal-oxide-semiconductor (CMOS)-compatible process is an ideal approach.

Recently, study of GeSn opens a new venue for light detection in a broad IR wavelength range [14-18]. As all-group-IV alloy, GeSn optoelectronics leverages the mature Si CMOS process and holds great promise for seamlessly integrating with Si electronic devices. By increasing Sn composition in GeSn alloy, detection coverage could be extended from shortwave-IR (SWIR) to mid-IR (MIR). Combining a GeSn absorption region with a Si multiplication layer in a SACM APD has been demonstrated [19-22]. Preliminary results have been reported showing the gain behavior. So far, almost all reported responsivities of GeSn APDs were measured at 1.55 μm or 1.3 μm to the best of our knowledge. Moreover, very few reports could demonstrate a systemic study including material, optical, and device characterizations, from which the processes of carrier generation, transportation, and multiplication leading to unambiguous avalanche gain can be in-depth analyzed. In this work, a GeSn on Si SACM APD grown using a chemical vapor deposition (CVD) reactor was demonstrated. A 240-nm-thick Ge buffer was employed to ease the lattice mismatch between Si and GeSn, which also enables the electron transportation from GeSn absorber to Si multiplication layer. The maximum responsivities were obtained as 0.55 A/W and 0.34 A/W under 1.55 μm and 1.9 μm excitation lasers, respectively. At 77 K, the peak gain of 3.44 was achieved.

## 2. Material growth and characterization

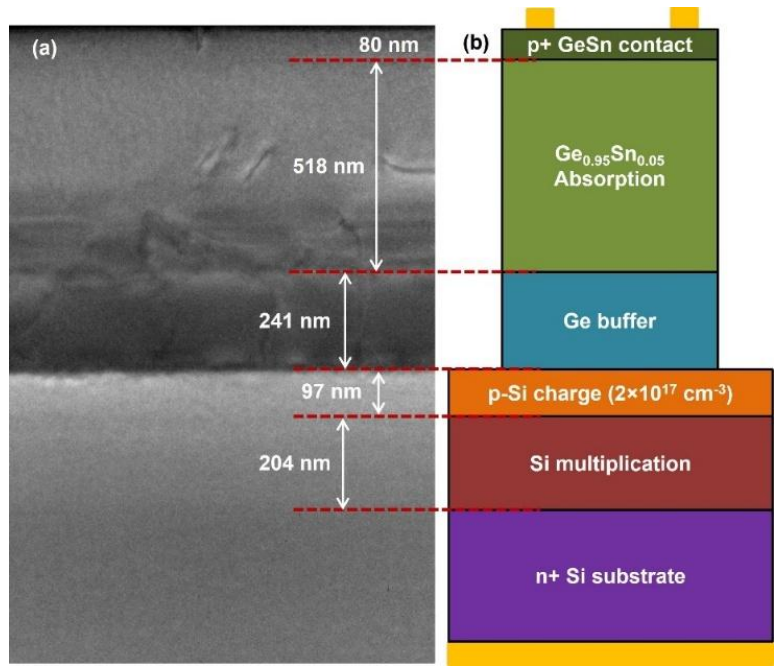

Figure 1. (a) TEM image of APD layer stack. Each layer can be clearly resolved. (b) schematic of APD structure.

The GeSn on Si SACM APD structure was epitaxially grown using an industrial standard ASM Epsilon® 2000 Plus reduced pressure chemical vapor deposition (RPCVD) reactor. The commercially available SiH4, GeH4, and SnCl4 were used as Si, Ge, and Sn precursors, respectively. The transmission electron microscopy (TEM) image in Fig. 1(a) clearly shows the layer stack, which includes (along the growth direction): i) a 204-nm-thick Si multiplication layer on Si substrate; ii) a 97-nm-thick Si charge layer; iii) a 241-nm-thick Ge buffer layer; iv) a 518-nm-thick $Ge_{0.95}Sn_{0.05}$ absorption layer; and v) a 80-nm-thick Ge0.95Sn0.05 contact layer at the very top. The interfaces between each two layers can be clearly resolved.

The Ge buffer layer was grown to accommodate the lattice mismatch between Si and GeSn. Based on previously reported GeSn APD, the Ge buffer design is one of determinative factors for device success. A thick Ge buffer could improve the GeSn material quality, but on the other hand it is unfavorable for photogenerated carrier transportation from GeSn absorber to Si multiplication layer. In this work, a non-trivial thin Ge buffer design was employed: while ensuring the high-quality GeSn absorber, the 240-nm-thick Ge buffer allows for the photo generated carriers to be effectively collected by multiplication layer, leading to the unambiguous avalanche gain. Note that when Ge buffer is thin, the surface strain is totally different from that of thick buffer, which greatly affects the following GeSn growth such as Sn incorporation.

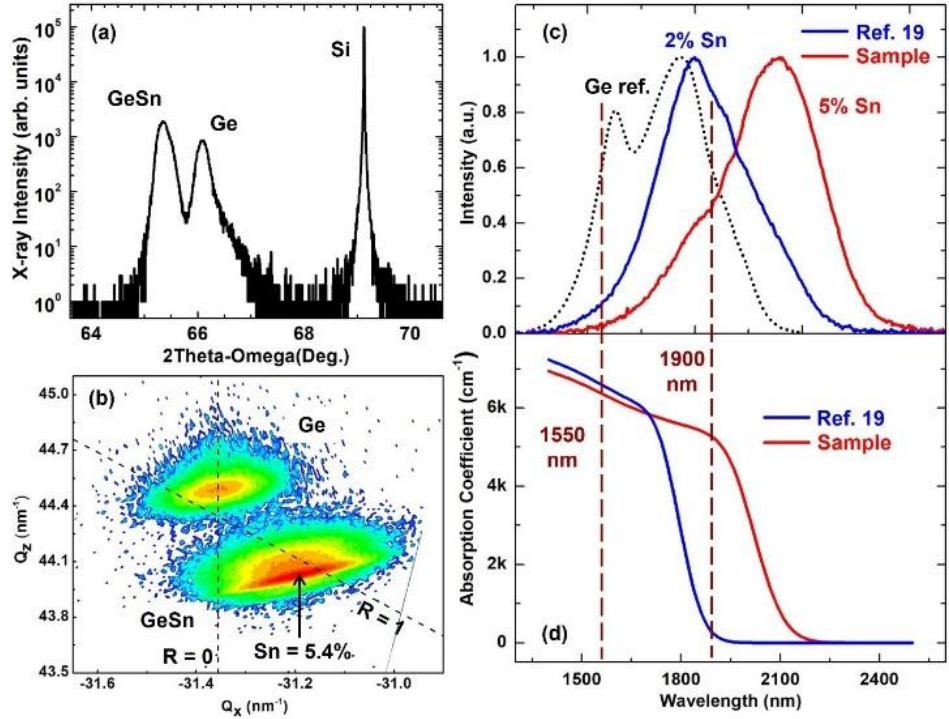

Figure 2. (a) XRD 2θ-Ω scan showing clearly resolved GeSn peak. (b) XRD RSM contour plot showing almost fully relaxed GeSn layer. (c) Room temperature PL spectral. The PL spectra of a Ge and a 2% Sn references are also plotted for comparison. (d) Spectroscopic ellipsometry showing the absorption edge beyond 2.1 μm.

X-ray diffraction (XRD) measurement was performed using a Panalytical X'Pert Pro Materials Research Diffractometer equipped with a 1.8 kW Cu Kα1 X-ray tube, a standard four-bounce Ge (220) monochromator, and a Pixel detector. Figure 2(a) shows the 2θ-Ω scan. The Si, Ge and GeSn peaks can be clearly resolved at 69°, 66.2° and 65.2°, respectively. The narrow peak width indicates a high material quality. The reciprocal space mapping (RSM) contour was plotted in Fig. 2(b), revealing that the GeSn layer is almost fully relaxed. The Sn composition was extracted as 5.4% via data fitting.

Photoluminescence (PL) spectroscopy was measured using a standard off-axis configuration with a lock-in technique to amplify the signal. A 1064 nm pulsed laser was used as the pumping source. The PL emission was collected by a grating-based spectrometer and then sent to an InSb detector with a cutoff wavelength of 5.0 μm. Figure 2(c) shows the normalized room temperature PL spectra of three samples: the GeSn sample with 5% Sn in this work, a Ge reference, and a GeSn sample with 2% Sn that was reported in ref. 19. As Sn composition increases, the PL peak shifts towards longer wavelength as expected. The obtained PL peak at 2180 nm unambiguously indicates that the emission is from GeSn absorber. Further optical characterization was carried out by employing spectroscopic ellipsometry technique [23]. A Variable-Angle Spectroscopic Ellipsometer (WVASE32) in the range from 1300 to 2500 nm was used to investigate the spectral absorption coefficient, as shown in Fig. 2(d). Compared to the reference sample with 2% Sn (cutoff at 1900 nm), the 5% Sn sample in this work exhibits longer wavelength cutoff beyond 2100 nm, matching with PL measurement results.

## 3.  GeSn APD device characterization

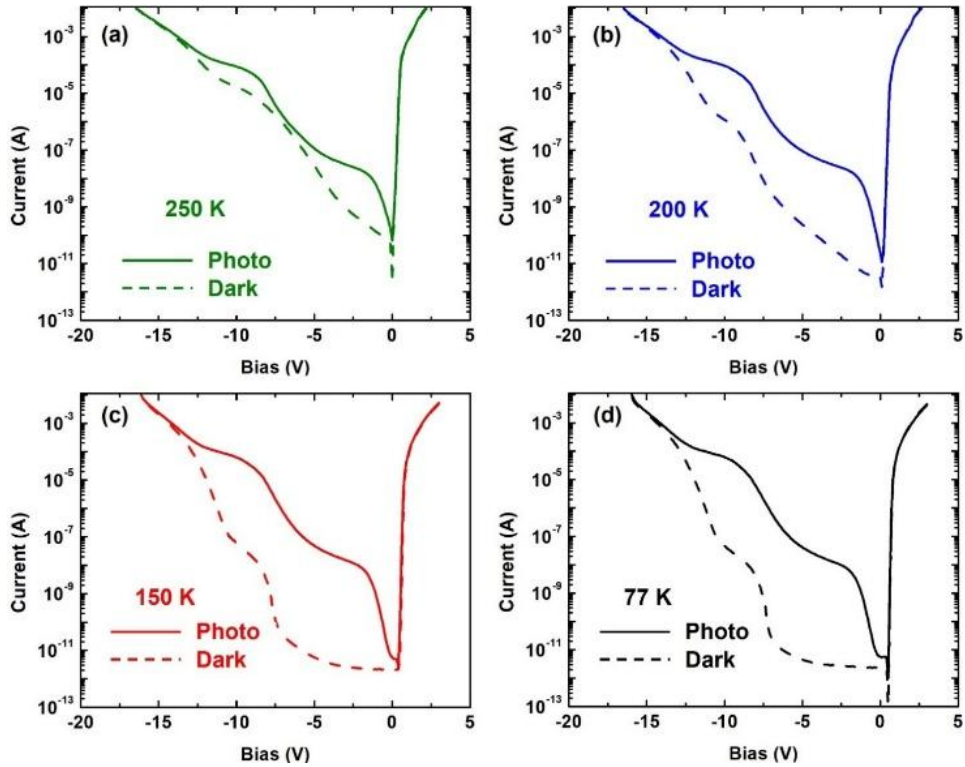

Figure 3. Measured dark (dashed curves) and photocurrents (solid curves) at (a) 250 K. (b) 200 K. (c) 150 K. (d) 77 K.

The APD devices with square mesa (side lengths of 750 μm) were fabricated using standard photolithography to define the patterns, followed by a wet chemical etching process using the mixture of HCl, $H_2O_2$, and DI $H_2O$ to form the mesa. The total etching depth is ~ 840 nm and automatically stops at Si charge layer. The 10/300 nm Cr/Au was used to form ohmic contact, which was deposited on top of GeSn mesa surface and at the backside of heavily doped Si substrate. The schematic cross-sectional view of the device is shown in Fig. 1(b). For temperature-dependent characterization, devices were wire bonded and mounted in an electrically isolated cryostat with $CaF_2$ windows to maximize IR light transmission.

Current-voltage (I-V) characteristics were measured using a Keysight B2902b source measurement unit (SMU). For photocurrent measurements a 1.55 μm laser with 0.5 mW power was used as light excitation. Figure 3 shows the dark and photo I-V curves at temperatures from 77 to 250 K. At each temperature, the typical rectifying characteristic was obtained under forward biasing. As temperature decreases, the dark current under reverse biasing decreases as expected. At 77 K, the dark current increases steeply at -8 and -10 V, corresponding to the onset and fully depletion of p-type Si charge layer and Ge buffer layer, respectively, as shown in Fig. 3(d). The depletion of Si charger layer allows for the carrier generated in Ge buffer transporting to Si multiplication layer, while the further depletion of Ge buffer layer enables the transport of carriers generated in GeSn absorber to Si multiplication layer. It is worth noting that the distinct photocurrent can be observed at 250 K, indicating the high-quality device.

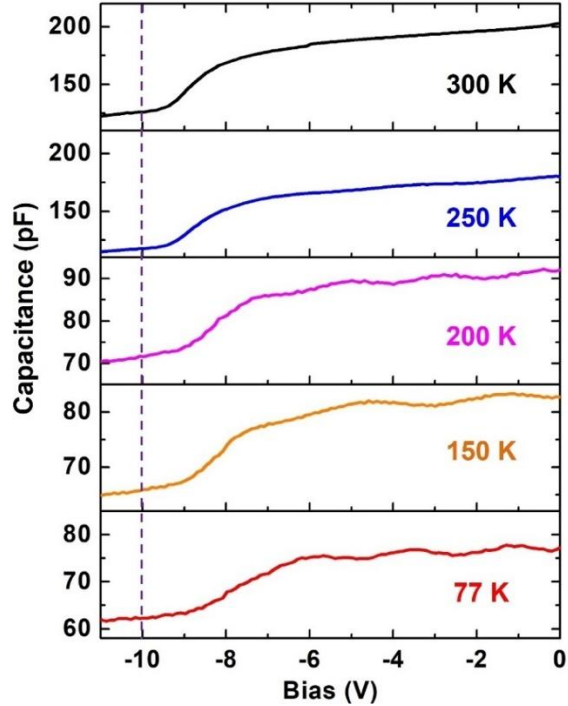

Figure 4. C-V characteristics at different temperatures. The plateau feature at -10 V was observed at each temperature.

The temperature-dependent capacitance-voltage (C-V) measurement was performed using a Keithley 590 CV Analyzer at frequency of 100 kHz. The larger capacitance at higher temperatures is likely due to a higher density of residual carriers in GeSn and Ge layers as a result of unintentional background doping, as shown in Fig. 4. At -11 V, due to steeply increased dark current, the measured C-V data are not reliable. At all temperatures, the C-V curves exhibit a plateau feature around -10 V, indicating a punch-through behavior, referring to that the depletion region extends through the Si charge layer to the edge of the Ge buffer layer. Note that at -10 V, the Si multiplication, Si charge layer, and Ge buffer layers are fully depleted, while the GeSn absorbing layer is partially depleted. This will be in-depth discussed in section 4.

Spectral response was measured using a Fourier transform infrared spectroscope (FTIR). Photo response was extracted across a 47 kΩ resister in series with the device using a pre-amplifier that fed into the FTIR. A 1300 nm long pass filter was employed to eliminate the absorption in the Si multiplication layer. Figure 5 shows the temperature-dependent spectral response measured at -12 V. At 77 K, the response cut off wavelength is at 1.93 μm. As temperature increases to 250 K, the response cut off shifts to 2.14 μm due to reduced bandgap energy. The shaded areas indicate the photo response beyond 2.0 μm at above 200 K. Moreover, the spectral response from previously reported APD with 2% Sn absorber was also plotted for comparison. Due to lower Sn composition, the wavelength cut off at 2.0 μm and 250 K was observed.

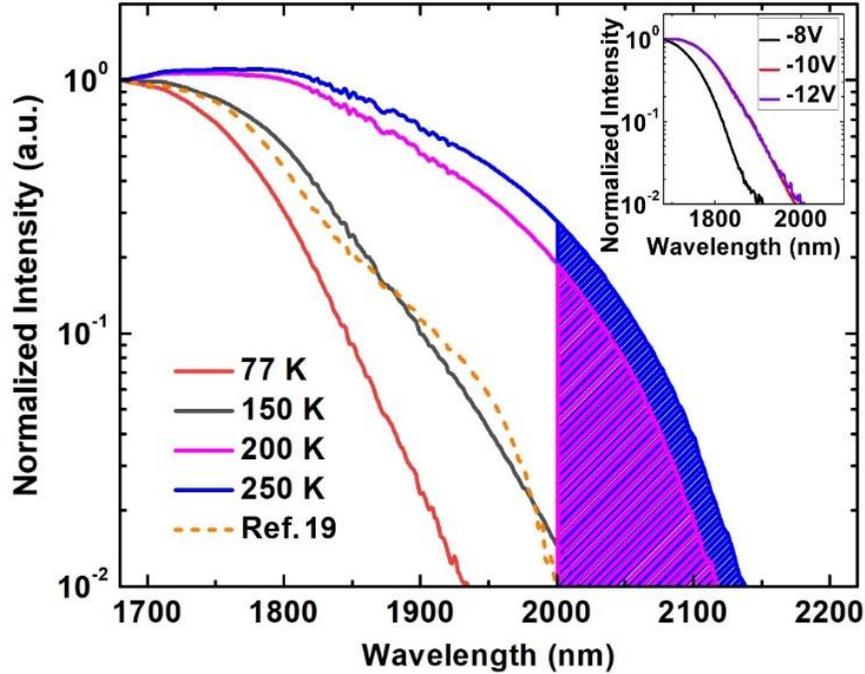

Figure 5. Temperature-dependent spectral response at -12 V. Shaded areas indicate the spectral response beyond 2.0 μm at 200 K and 250 K. The dashed curve shows response cut-off at 2.0 μm at 250 K in reference 19. Inset: spectral response at different bias voltages showing clear shift at -10 V.

The bias-dependent spectral response is shown in Fig. 5 inset. As the reverse bias voltage increases from 8 V to 10 V, the wavelength cut off exhibits an obvious shift from 1.9 μm to 2.0 μm. This abrupt jumping phenomenon can be explained as follows: at -8 V, the Ge buffer layer is partially depleted, and the spectral response mainly corresponds to Ge absorption. At -10 V, since the Ge buffer layer is fully depleted and punch-through occurs, the GeSn absorption contributes to spectral response, leading to that the response cut off shifts towards longer wavelength. As reverse bias voltage further increases, the cut off wavelength remains at 2.0 μm. The bias-dependent spectral response confirms the punch-through voltage at -10 V that was observed in C-V measurement.

For responsivity measurement, a 1.55 μm and a 1.9 μm lasers were utilized as excitation sources with 0.5 mW power. The signal was chopped at a frequency of 377 Hz and measured across a series resistor of 15 Ω, then fed into a lock-in amplifier. Responsivity was then calculated from photocurrent flowing through the resistor. Figure 6(a) shows measured responsivity under 1.55 μm excitation laser at different temperatures. Generally, a higher responsivity is obtained at higher temperature due to reduced bandgap energy which leads to enhanced light absorption. At lower temperature (< 150 K), responsivity starts to increase dramatically at -8 V corresponding to onset of Ge buffer depletion. On the other hand, as temperature increases, the responsivity increasing rate decreases. This is mainly because of the decreased impact ionization rate in Si multiplication region at higher temperature. Under 1.55 μm excitation, both Ge buffer and GeSn absorption layers could contribute to the responsivity. The maximum responsivity was measured as 0.55 A/W at 250 K.

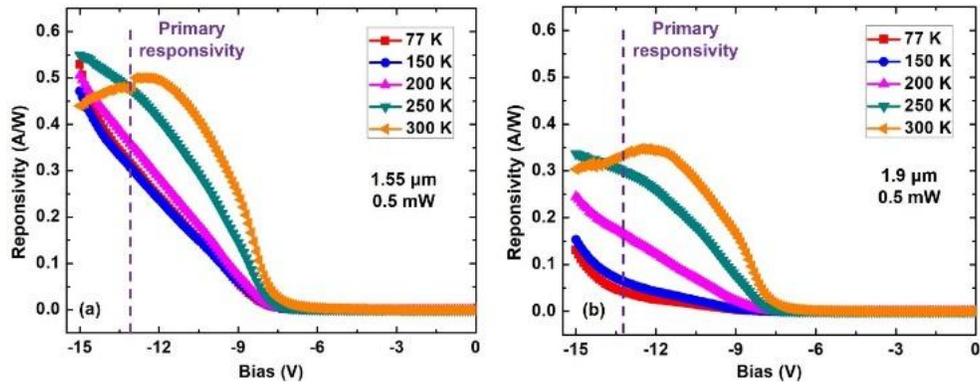

Figure 6. Temperature-dependent responsivity measured under (a) 1.55 μm laser and (b) 1.9 μm laser. The primary responsivity was conservatively selected at - 13 V.

Responsivity curves under 1.9 μm excitation laser are plotted in Fig. 6(b). Compared to under 1.55 μm laser, the values of responsivity decrease at each temperature as a result of that only GeSn layer absorption contributes to photo response. Note that at the temperatures below 150 K, responsivity starts to show significant increase at -10 V, corresponding to the fully depletion of Ge buffer layer and onset depletion of GeSn absorption layer. The maximum responsivity was measured as 0.34 A/W at 250 K.

## 4. Further analysis on device operation

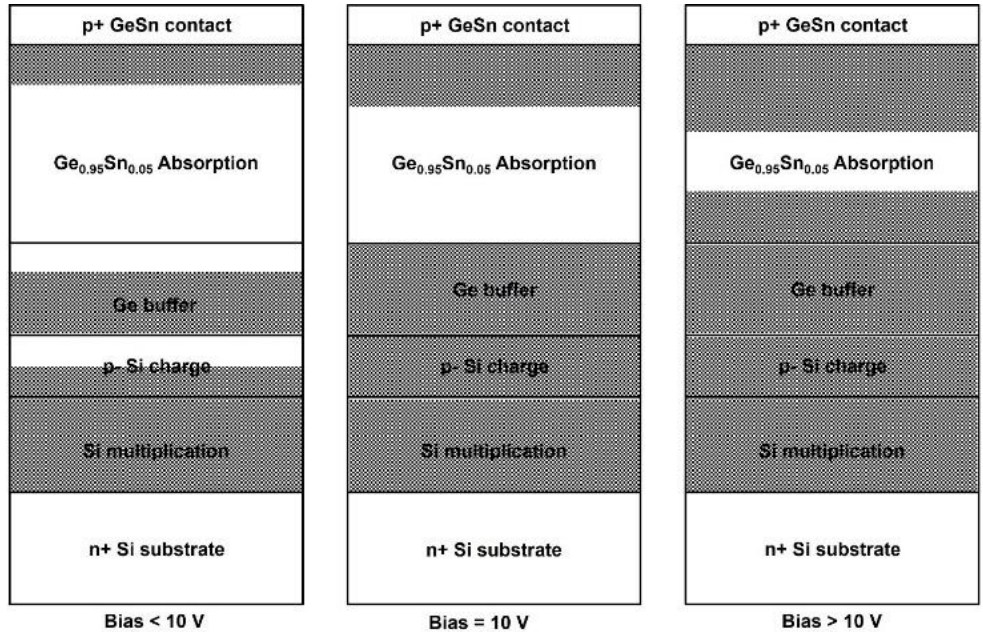

Figure 7. Progressive depletion region at different reverse bias voltages.

The activation energies were extracted from dark current shown in Fig. 3 to understand the depletion process. At reverse bias voltage of 5.5 V ~ 8 V, the activation energy of 0.541 eV was obtained, clearly indicating that the dark current is determined by Si layers (Fig. 7 left). At -10 V, the activation energy drops to 0.309 eV, which is close to the half of bandgap energies of Ge buffer (0.66 eV) and GeSn absorber (0.59 eV) but closer to that of GeSn. Considering the bandgap increase of Ge at lower temperature, this activation energy suggests the onset of GeSn layer depletion. At this voltage the Ge buffer and Si charge layers are fully depleted (Fig. 7 middle), and the connection of Ge and Si depletion regions implies the punch-through occurrence, which is also supported by C-V plateau (Fig. 4) and spectral response sudden red-shift (Fig. 5 inset) at -10 V. At reverse bias voltage of 12 V ~ 14 V, the activation energy remains of 0.308 eV, meaning that the GeSn layer is still partially depleted (Fig. 7 right). At 15~16 V, the activation energy further drops to 0.147 eV. This may be due to that the avalanche process has been already started, and the dark current is dominated by defect-assisted tunnelling under high electric field in GeSn. Note that for this device, the avalanche breakdown takes place before full depletion of GeSn absorber. This is not ideal for APD operation. However, owing to the non-trivial Ge buffer design, the high quality GeSn absorber allows for the photo generated electrons transporting to Si multiplication layer even with partial depletion, and therefore the avalanche gain was obtained.

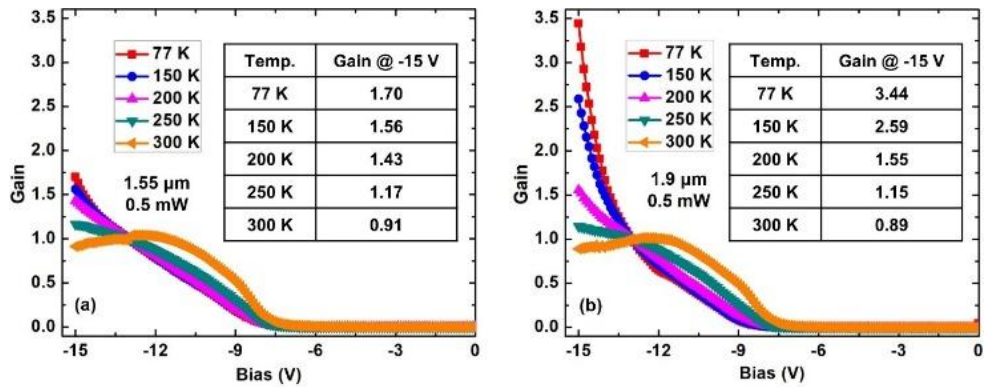

Figure 8. Calculated avalanche gain under (a) 1.55 μm and (b) 1.9 μm laser.

To calculate the avalanche gain, the primary responsivity needs to be determined. Considering that the GeSn layer is not fully depleted at punch-through voltage of -10 V, and the activation energy is close to GeSn half bandgap at -10 V ~ -14 V, we conservatively selected the values at -13 V as primary responsivity, which are 0.30 A/W and 0.05 A/W at 77 K under 1.55 μm and 1.9 μm excitation lasers, respectively (Fig. 6). Figure 8(a) shows the calculated gain under 1.55 μm excitation laser at different temperatures. The maximum gain of 1.70 was obtained at 77 K and -15 V. As temperature increases, the gain decreases as a result of reduced impact ionization rate. Note that ever at 250 K, the gain of 1.17 is greater than unity. Figure 8(b) shows the calculated gain under 1.9 μm excitation laser at different temperatures. The maximum gain of 3.44 was obtained at 77 K and -15 V. The measured gain under 1.9 μm laser is higher than that under 1.55 μm laser at most temperatures. This can be interpreted by the Franz–Keldysh effect, which enhances the 1.9 μm light absorption at high electric field, leading to more rapidly increased responsivity, and consequently the higher gain.

It is well acknowledged that for ideal APD operation, the full depletion of absorber should take place at lower reverse bias voltage than breakdown. The simulation of electric field distribution was performed to illustrate the GeSn depletion and breakdown voltages, as shown in Fig. 9. It is clear that the intrinsic doping (unintentional background doping) level in GeSn absorber and Ge buffer plays an important role. At doping concentration of $5\times10^{16}$ cm$^{-3}$, the breakdown voltage is ~17 V, while the full depletion of GeSn occurs at greater than 20 V, i.e., the device would work but will always operate in non-ideal situation. If the doping concentration could reduce to $2\times10^{16}$ cm$^{-3}$, the breakdown voltage increases to greater than 20 V while the full depletion voltage of GeSn dramatically reduces to ~14 V, shifting the device to ideal operating mode.

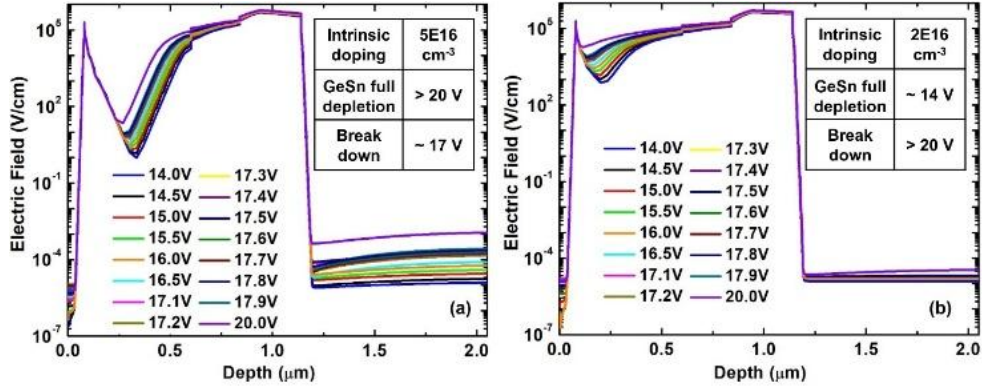

Figure 9. Electric field distribution at different reverse bias voltages with the intrinsic doping concentrations of (a) $5\times10^{16}$ cm$^{-3}$ and (b) $2\times10^{16}$ cm$^{-3}$ in GeSn absorber.

Generally speaking, the higher Sn incorporation, the higher intrinsic doping concentration. To broaden the IR wavelength coverage of GeSn APD, the higher Sn composition absorber is desired. This requires the advanced material growth technique with optimized Ge buffer design to simultaneously offer high material quality and high carrier collection efficiency. In addition, to further improve the device performance, the following design will be considered for the future GeSn APD development: i) increase the thickness of Si multiplication layer. The measured thickness of Si multiplication layer is 204 nm in this work. According to reported high-performance Si APD, a 500-nm-thick layer is a viable design. A thicker Si multiplication layer could raise the breakdown voltage. ii) decrease the thickness of GeSn absorber. With high-quality material, a thinner GeSn absorption layer of 300~400 nm would not only be sufficient for carrier generation but also facilitate the full depletion at relatively lower voltage. It is worth noting that with increased Sn composition to unlock the longer wavelength operation, the Ge buffer design should be modified accordingly to enable a low intrinsic doping level and thin GeSn layer serving as an excellent absorber.

## 5. Conclusion

In summary, we have demonstrated a Si-based SACM APD using a GeSn with 5% Sn as absorber and a Si as multiplication layer. A 240-nm-thick Ge buffer layer was employed to accommodate the lattice mismatch. This design also allows for the transportation of photo generated electrons from GeSn absorber to Si multiplication region. Spectral response is extended to 2.14 μm at room temperature. Dark I-V and C-V measurements indicated a punch-through voltage of -10 V, at which the Ge buffer and Si charge layers are fully depleted, and their depletion regions are connected, while the GeSn absorption layer starts to deplete. The primary responsivities at 77 K were measured as 0.30 A/W and 0.05 A/W under 1.55 μm and 1.9 μm excitation lasers, respectively. With partially depleted GeSn absorber, the calculated maximum gain values were 1.70 and 3.44 at 77 K under 1.55 μm and 1.9 μm lasers, respectively. Even at 250 K, the avalanche gain was obtained greater than unity. The simulation of electric field distribution indicated that a lower intrinsic doping level in GeSn absorber and Ge buffer would facilitate the full depletion and make it occur at a lower voltage than breakdown, which is desirable for APD operation.

**Funding.** Air Force Research Laboratory (AFRL)/AFWERX (Contract No. FA864922P0744).

**Acknowledgment.** Shui-Qing Yu would like to thank Northrop Grumman Corporation (NGC) for gift funding to partially support this work. He also gives his thanks for many fruitful discussions and encouragements from Drs. Alex Toulouse and Leye Aina at NGC for this work and greatly appreciates their input.

**Disclosures.** The authors declare no conflicts of interest.

**Data availability.** Data underlying the results presented in this paper are not publicly available at this time but may be obtained from the authors upon reasonable request.